**Multi-confocal fluorescence correlation spectroscopy**

Remi Galland[1], Jie Gao[1], Meike Kloster[1], Gaetan Herbomel[2], Olivier Destaing[3], Martial Balland[1], Catherine Souchier[2], Yves Usson[4], Jacques Derouard[1], Irene Wang[1], Antoine Delon[1]

[1]Universite de Grenoble 1, CNRS, Laboratoire de Spectrometrie Physique UMR 5588, BP 87, 38402 Saint Martin d'Heres, France, [2]Universite de Grenoble 1/ INSERM, Institut Albert Bonniot, U823, equipe 10, Stress et DyOGen, La Tronche, BP 170, 38042 Grenoble Cedex 9, France, [3]Universite de Grenoble 1/ INSERM, Institut Albert Bonniot, U823, equipe DySAD ERL CNRS 3148, La Tronche, BP 170, 38042 Grenoble Cedex 09, France, [4]Universite de Grenoble 1, Institut d'Ingenierie et de l'Information de Sante, Laboratoire TIMC, La Tronche, 38706 La Tronche cedex, France

TABLE OF CONTENTS

**1. ABSTRACT**

We report a multi-confocal Fluorescence Correlation Spectroscopy (mFCS) technique that combines a Spatial Light Modulator (SLM), with an Electron Multiplying-CCD camera (EM-CCD). The SLM is used to produce a series of laser spots, while the pixels of the EM-CCD play the roles of virtual pinholes. The phase map addressed to the SLM, calculated by using the spherical wave approximation, makes it possible to produce several diffraction limited laser spots. The fastest acquisition mode leads to a time resolution of 100 µs. By using solutions of sulforhodamine G we demonstrated that the observation volumes are similar to that of a standard confocal set-up. mFCS experiments have also been conducted on two stable cell lines: mouse embryonic fibroblasts expressing eGFP-actin and H1299 cells expressing the heat shock factor fusion protein HSF1-eGFP. In the first case we could recover the diffusion constant of G-actin within the cytoplasm, although we were also sensitive to interactions with F-actin. Concerning HSF1, we could clearly observe the modifications of the number of molecules and of the HSF1 dynamics during heat shock.

**2. INTRODUCTION**

The cellular environment is a highly heterogeneous and dynamic medium, which exhibits fast spatial and temporal changes. For this reason, it is important to develop Fluorescence Correlation Spectroscopy (FCS) methods that enable simultaneous measurements at different locations within a cell. Not only would such methods provide more complete information about the cellular machinery, but also, by performing a large number of measurements in parallel, one would obtain statistically significant results and assess cellular variability in a more time-efficient way.

Among the different approaches one can distinguish three families: Image Correlation Spectroscopy, that exploits the information implicitly imbedded in images recorded with conventional means; scanning FCS, where auto-correlation curves are constructed with repeated scans of the same 1D zone (usually linear or circular); multipoint or multi-confocal FCS techniques, where at least two laser spots or two point detectors are used.

Image Correlation Spectroscopy (ICS), the oldest technique, was proposed by Petersen (1). In its initial version, this technique deals with immobile molecules or particles, the number of which (in the PSF volume) being inversely proportional to the amplitude of the image (*i.e.* spatial) autocorrelation function. Any biochemical phenomenon that affects the concentration, such as aggregation, is then monitored through the autocorrelation amplitude. If, nevertheless, the molecules or particles are mobile, then information about mobility can be retrieved by recording a stack of images (providing the acquisition rate is high enough). Wiseman (2, 3) developed this approach to produce cellular maps of molecular densities, interactions, diffusion rates, and the net direction and magnitude of non-random, concerted molecular movement. The local diffusion constants (or velocities) can thus be determined by fitting the temporal autocorrelation functions, averaged over the relevant regions. The advantage of ICS is that, since entire images are recorded, it is straightforward to choose regions of interest (ROI) for the analysis. In its more recent developments (4, 5), the time resolution of this kind of methods can reach that of single point measurements (6), the price to pay being a low spatial resolution. Note that the scan modality is given by the raster-scan mode of the confocal microscope.

A more flexible method consists in scanning the region of interest during FCS acquisition, but along a chosen laser trajectory (7, 8). This technique, so called scanning FCS (sFCS), has been applied with various modalities (9-13). However, here also there is a compromise between the temporal resolution, the spatial resolution and spatial extension, since only one laser spot and one single point detector are used. One cannot get independent measurements at different points arbitrary located in the field of view, with the temporal resolution of single point FCS.

To overcome this barrier, it is necessary to develop multiple laser spot systems. Several techniques have already been used to produce separate laser spots. In one of the pioneering experiments, based on elastic scattering of light by latex spheres, two volumes were obtained with two different $Ar^+$ laser lines (458 and 514 nm) (14). A few years later, two papers reported spatial cross-correlation spectroscopy under confocal geometry (15, 16). However, all these experiments used polarizing beam splitters and/or Wollaston prisms to split and recombine the two, almost parallel, laser beams (15, 17). In this case, it is not possible to adjust the distance between the two observation volumes. Nevertheless, a Wollaston prism is an interesting optical component because it can be easily inserted in the illumination path of commercial FCS systems, providing a fixed and known distance between the two foci which can therefore be used as an external ruler (17, 18). Passive Diffractive Optical Elements (DOE) have also been used to produce gaussian foci of sub-micrometer diameter and perform FCS and spatial cross-correlation spectroscopy experiments (19, 20). However, here also there is some lack of flexibility, since each DOE is fabricated once and for all by electron beam lithography. Promising ways to flexibly address several laser spots within the ROI are spinning disk confocal microscopes (21, 22) and dual head confocal microscopes (23). In the former case, while a dramatic increase of the number of available independent locations (up to ~ $10^4$) is possible, one nevertheless encounters the limitations of spinning disk systems, in terms of limited temporal resolution. In comparison, the dual laser spot scanning confocal system recently developed is a versatile instrument, since it makes it possible to either, park two laser spots anywhere in the field of view, with the ultimate temporal resolution of a standard FCS microscope, or to perform sFCS acquisitions.

However, none of the proposed techniques can provide fast simultaneous measurements on an arbitrary number of confocal volumes that can be located independently. Therefore we believe that there is still a need for a multi-confocal FCS (hereafter called mFCS) system for living cell studies, which would involve:
 i) a flexible way to address, simultaneously, the desired laser spots at various locations within the biological medium;
 ii) a matrix of fast, point-like detectors.

Concerning the excitation side, Spatial Light Modulators (SLM) are now used for microscopy applications, mainly for optical tweezers (24) and adaptive optics, since these devices make it possible to control the laser illumination geometry. We have recently demonstrated the potential of SLM for mFCS by measuring, at the single particle level, both active transport (*i.e.* a flow) and passive transport (in that case, permeability through a phospholipidic membrane) (25). Concerning the detection side, Electron-Multiplying CCD (EM-CCD) cameras are especially promising, since each pixel is a single photon point-like detector (26-28) and because the on-chip amplification makes it compatible with a fast read-out rate (10 MHz) and a high signal to noise ratio.

Here, we report the first results obtained with a mFCS system built by combining a SLM to address several points of interest and a EM-CCD camera for fast recording of the corresponding photon streams in a confocal mode. After performing control experiments using dye solutions, we have demonstrated the potential of our system for living cell measurements by addressing two biological issues.

First, we focused on the actin skeleton of fibroblast cells (29). Living cells are characterized by their ability to adapt to their physical environment by modulating their mechanical properties. These properties are mainly attributed to the dynamical properties of cell actin cytoskeleton and to a large number of associated proteins that will affect actin assembly and dynamics. Basically, actin cytoskeleton is a complex three-dimensional network of protein filaments. Those filaments result from the polymerization of monomers of globular actin (G-actin). Actin filaments are permanently submitted to several cycles of polymerization/depolymerization depending on polymer polarity and actin monomer dynamics. The nucleation and/or branching of new actin filaments and their assembly are critical for the formation of sub-cellular functional structures such as filopodia or lamellipodia. Because actin dynamics implies slow processes, such as filament turnover and detachment rate constants at the filament ends (30), Fluorescence Recovery After Photobleaching (FRAP) or PhotoActivation of Fluorescence (PAF) are generally preferred (31). In order to characterize and quantify these processes with higher temporal resolution, it would be important to obtain spatially resolved molecular concentration maps for the different constituents of the cytoskeleton machinery. Up to now, only one FCS study has been devoted to the analysis of filaments dynamics and it did not provide any information about G-actin mobility (32).

The other biological question we have addressed is that of the nuclear response of cells to a thermal stress. More precisely, we are interested in the dynamics of the transcription factor HSF1 (Heat Shock Factor 1) that relocalizes, after heat shock, in a limited number of nuclear structures termed nuclear stress bodies (NSB). The exposure of a cell to a stress, such as a heat shock, induces a transient and ubiquitous response, called the "heat shock response", the function of which is to protect the cells against deleterious effects potentially induced by stressful conditions. This response is associated with important changes in gene expression, among which are those of heat shock proteins which are controlled by the transcription factor HSF1 (33). Potentially our mFCS technique should be particularly interesting when studying highly structured and inhomogeneous biological systems (such as the actin cytoskeleton), or when the system is subject to temporal variations (such as the heat shock response).

In the following, we first present the experimental device in the Material and methods section. To characterize the system, mFCS measurements were performed with fluorescent dye solutions, as reported in the Experimental results section. Then, the first results on living cells are presented. Finally, in the Discussion and conclusion section, we summarize the performances and limitations of our mFCS setup and suggest a number of improvements, which should enhance the potential of this method for biological applications.

## 3. MATERIAL AND METHODS

### 3.1. Optical setup

The FCS measurements were performed on a homebuilt experimental setup as shown on Figure 1. This setup is based on an inverted microscope body (Olympus IX70). Fluorescence excitation is performed using a solid-state laser emitting 20mW of 488 nm-wavelength continuous-wave light (85-BCD-020, Melles Griot). The beam is sent into a single-mode fiber in order to spatially clean its wavefront. Then it is directed to the spatial light modulator (SLM) composed of 800×600 20 µm-pixels (LCOS-SLM X10468-01, Hamamatsu). It is worthwhile to mention that the driver of the SLM converts the input digital signal (from the DVI receiver) to an analog voltage. With this addressing method (as opposed to digital addressing), the phase pattern is stable in time and does not show short time fluctuations. Before reaching the SLM, the beam is expanded using a telescope to cover the active area of the SLM, and goes through a half-wave plate and a polarizer since the SLM only diffracts horizontally polarized radiation. When the adequate phase map is addressed to the SLM, the beam is focused in a plane which is optically conjugate with the microscope object plane. The beam is then collimated before being reflected by a dichroic mirror (PB 505, Olympus) toward an oil immersion objective lens (×60, NA=1.4, Olympus). This collimation lens is also used to image the SLM onto the back aperture of the objective lens to ensure that all the diffracted light is coupled into the objective. The laser power entering the objective lens is varied between 10 and 50 µW (according to the total number of spots and the propensity of the sample to photobleach)

Fluorescence emitted by the sample is spectrally filtered (PH 510, Olympus) then directed to a side port of the microscope where it passes through a pair of achromats producing a magnification of ×5.3, which involves a total magnification of about ×316. Then the collected fluorescence is sent to one of two detection paths. The first leads to a multimode fiber of 100 µm core diameter connected to an avalanche photodiode (SPCM-AQR-13, Perkin Elmer) to perform standard FCS measurements. The fiber core acts then as a pinhole to ensure confocal detection. On the second path, another pair of achromatic lenses demagnifies the fluorescence image by a factor 4 before sending it on an electron-multiplying CCD camera (iXon, Andor Technology). In this case, each pixel of the camera acts as an individual pinhole for parallel confocal detection. A mask can be positioned before the lenses in a plane conjugate with the microscope object plane to prevent out-of-focus light from falling on unused parts of the CCD chip, which is required when using special readout modes of the EM-CCD camera. This mask consists of a transparent pellicle film with a black printed pattern or a razor edge if only one side of the captor has to be masked.

### 3.2. Phase map calculation

To create an arbitrary set of diffraction limited spots in the microscope object plane, we displayed on the SLM the corresponding phase map computed by a spherical wave and superposition approach (34). As described in the previous paragraph, we want the SLM to focus the incoming plane wave into a number of freely chosen spots at a finite distance. The principle of the algorithm is simple: we assume that each desired spot results from a converging spherical wave. Then the phase function is obtained from the back-propagation and superposition of the spherical wavefronts in the plane of the SLM.

We define $r_m(x,y)$ as the distance between the pixel of the SLM at coordinates $x,y$ and the $m^{th}$ spot (Figure 2). In our case, all the spots are created in the same plane on the z-axis, optically conjugate with the microscope object plane. This plane, which can be considered as the SLM focal plane, is at a distance $f$ (we used $f$ = 364 mm throughout this work) from the SLM. It should be noted that off-focus spots could be obtained exactly in the same way.

$$r_m(x, y) = \sqrt{(x - x_m)^2 + (y - y_m)^2 + f^2} \quad (1)$$

To generate several spots, the complex amplitude of the electric field immediately after the SLM is given by the sum of spherical waves converging at different spots:

$$u_{out}(x, y) = \sum_m a_m \frac{\exp\left(-i\frac{2\pi}{\lambda} r_m(x, y)\right)}{r_m(x, y)} \quad (2)$$

where $a_m$ are amplitude factors, that take into account the obliquity factor, *i.e.* $\cos\psi_m(x, y) = f / r_m(x, y)$.

It should be noted that this exact field cannot be created using the SLM since the latter can only modulate the phase of the incident wave and not its intensity. In a first approximation, we ignore the intensity mismatch and apply the following phase pattern to the SLM:

$$\phi(x, y) = \arg\left(\sum_m a_m \frac{\exp\left[-i\frac{2\pi}{\lambda} r_m(x, y) + \theta_m\right]}{r_m(x, y)}\right) \quad (3)$$

Note that we add random phases $\theta_m$ (uniformly distributed in $[0, 2\pi]$) to each spot in order to reduce unwanted interferences leading to the occurrence of ghost spots, which deviate part of the laser energy from useful spots.

This 'spherical random superposition' algorithm has the advantage of short computational times. Diffraction-limited spots can be generated with reasonably good efficiency. However, the laser power is not distributed uniformly between different spots, as we will see in the Multiple spot measurement section (4.1.2). To improve the uniformity, it should be possible to use an iterative algorithm similar to the Gerchberg-Saxton algorithm (35) adapted to the spherical geometry, but at the cost of a slower computational speed.

### 3.3. Data acquisition and treatment
### 3.3.1. Data acquisition

The avalanche photodiode signal is treated by a homemade data acquisition system and correlated by software. The EM-CCD images are acquired with the Solis software supplied by Andor.

The maximum frame rate of the EM-CCD camera depends on the number of pixels read and the readout mode. Throughout this work, we used three distinct readout modes: 1) the standard frame transfer mode in which we define a subregion, called 'ROI mode'; 2) the 'Crop mode' in which only the bottom right corner of the CCD chip is read; 3) the 'Multi-track mode' in which one can choose a number of isolated rows at different positions on the chip and read only the signal falling on these rows. For the ROI mode, the frame rate decreases with the number of rows in the subregion and its distance to the bottom of the chip. For a subregion of 3 rows at the bottom of the chip, the maximum frame rate is 5000 Hz, which corresponds to a time resolution of 200 µs. The Crop mode exhibits faster readout with a frame rate of 9930 Hz (100 µs time resolution) for 3 rows. Changing the width of the crop area does not speed up the acquisition rate, so we always read entire rows. The useful rows must be at the bottom of the chip. In this mode, care should be taken to block light falling on the upper part of the camera since this would corrupt the useful signal. That is why a mask (in the present case, it is simply a razor edge) is placed at a position optically conjugated with the camera chip. Note that, with ROI and Crop modes, sufficient time resolution can only be obtained if the useful spots are all gathered within a limited number of rows. Therefore, the spots are typically aligned in a horizontal row and we cannot use the whole field of view of the camera. In this context, the Multi-track mode offers an interesting possibility since one can choose to read 'tracks' (consisting of one or several adjacent rows) at arbitrary vertical positions on the CCD chip. In this way, information can be obtained from different vertical positions all over the field of the camera by creating spots at the height of the pixel rows to be read. The maximum frame rate is 4348 Hz for reading 3 rows (one in the upper part, one in the middle and one in the bottom part of the chip).

For all the measurements presented here, the vertical shift speed was set to 0.1 µs per line and the 'baseline clamp' option, which corrects for electronic offsets, was not chosen in order to speed up the readout. As we will explain below, offsets are corrected afterwards by subtracting dark measurements from each acquisition. The electron multiplying gain was set between 300 and 1000, which is more than sufficient for single photon detection. Each acquisition lasted typically 10 seconds; the corresponding number of images would vary depending on the frame rate. Every 5 acquisitions, a dark measurement of the same duration was taken by blocking the excitation light, to record the electronic offset.

### 3.3.2. Data treatment

Image sequences are treated offline and autocorrelation functions are calculated using Matlab (MathWorks). To calculate the autocorrelation function, the position of the brightest pixel (for each excitation spot) is determined by averaging all the images from one acquisition. The signal of this pixel is extracted as a function of time. The level of the pixel at the corresponding position in the dark acquisition is also extracted. The dark level generally exhibits a drift during the acquisition. To correct for this offset, the dark level as a function of time is smoothed (convolution with a one second-time window) and then subtracted to the time trace. Once this offset subtraction is done, an additional correction is performed to avoid distortions of the autocorrelation curves due to photobleaching (although it is limited to a few %) and variations due to cell motility and deformation. For this purpose, the offset-corrected time trace is divided by a smoothed time trace (obtained from the first one by calculating its convolution with a one second-time window). As a consequence, fluctuations on a time scale longer than 0.5 s cannot be recovered within the autocorrelation function. Then the normalized autocorrelation function of the time trace is calculated. The time interval between data points in the autocorrelation curve is made to increase almost exponentially with increasing delays. In the following, all the autocorrelation curves are averaged over at least 5 acquisitions and, to assess the reproducibility of the measurement, standard errors of the mean (SEM) are calculated for each data point.

Data plotting and nonlinear least-square fitting are performed with Origin (OriginLab) using the Levenberg-Marquardt algorithm and the standard diffusion model for one or two species. If we consider one freely diffusing species, the autocorrelation function is described by:

$$G(\tau) = 1 + \frac{1}{N} \left[ \left(1 + \frac{\tau}{\tau_D}\right) \sqrt{1 + \frac{\tau}{S^2 \tau_D}} \right]^{-1} \quad (4)$$

where $N$ is the average molecule number, $\tau_D$ is the characteristic diffusion time, $S$ is the structure parameter ($S=z_0/r_0$ where $z_0$ and $r_0$ are the axial and lateral $1/e^2$ radii of the effective volume). The diffusion constant $D$ can be obtained by $D = r_0^2 / 4\tau_D$.

Measurements in living cells could generally not be fitted with such a simple model, because fluorescent species show a distribution of mobilities, due to variations in size or interaction with their environment. In this work, autocorrelation curves obtained in cells are adjusted with a two component diffusion model:

$$G(\tau) = 1 + \frac{1}{(N_1 + N_2)^2} \left( N_1 \left[ \left(1 + \frac{\tau}{\tau_{D1}}\right) \sqrt{1 + \frac{\tau}{S^2 \tau_{D1}}} \right]^{-1} + N_2 \left[ \left(1 + \frac{\tau}{\tau_{D2}}\right) \sqrt{1 + \frac{\tau}{S^2 \tau_{D2}}} \right]^{-1} \right) \quad (5)$$

We believe that, in the present case, the second term does not refer to another diffusing species, but rather to interaction processes that hinders free diffusion.

### 3.4. Sample preparation
#### 3.4.1. Dye solution

FCS measurements in solution were performed using sulforhodamine G (SRG) molecules (Radiant Dyes, Wermelskirchen). These molecules were diluted to a final concentration of about 10 nM, either in pure water, or in a 95:5 glycerol-water mixture when a relatively long diffusion time was required (*e.g.* to test the EM-CCD camera). Since the viscosity of glycerol is very sensitive to temperature, we have performed all measurements at 20°C. The temperature was regulated using an objective heater and Delta T system from Bioptechs (Butler, PA). It is interesting to note that the oil immersion objective ensured a more uniform temperature throughout the sample than a water immersion one.

#### 3.4.2 Cell culture and reagents

GFP-actin (Clontech) was cloned into the retroviral vector pBabe (Addgene, Cambrige, USA). Viral particles were produced by transfecting pBabe-GFPactin into a packaging cell line, Phoenix ecotrope, with Lipofectamine 2000 according to the manufacturer protocol (Invitrogen). Their conditioned supernatant containing viral particles is used to infect primary mouse embryonic fibroblast (MEF). Cells were cultured at 37°C, 5% $CO_2$ in DMEM (Gibco BRL) with 10% (vol/vol) fetal calf serum (FCS, Sigma-Aldrich) and 100U/ml each of penicillin and streptomycin.

Stable H1299 cell lines expressing human Heat Shock factor 1 fused to eGFP were established and grew at 37°C, 5% $CO_2$ in RPMI 1640, 10% FBS (Lonza). A day before observation, cells were plated on culture dishes (Bioptechs, Butler, PA) and were then observed in 20 mM Hepes buffered HBSS + $Ca^{++}$ + $Mg^{++}$.

### 3.5. Heat shock treatment

To perform heat shock, we used an objective heater and a Delta T system from Bioptechs. Heat shock was induced by raising the temperature in the culture dish from 37°C to 43°C (the objective was heated only to 40°C to avoid damage). After an initial period of about 8 min, which was required to achieve the requested temperature, the cells were kept at the elevated temperature for about 35 min. mFCS measurements were performed before and after heat shock.

## 4. EXPERIMENTAL RESULTS

### 4.1. Measurements in solution
#### 4.1.1. Single spot measurements

Quantitative FCS measurements require an effective volume of well-known size. This volume depends on the excitation beam as well as the detection pinhole. In our setup, as opposed to a conventional confocal FCS experiment, the excitation spots are created using a spatial light modulator, and on the detection side, the pinhole is replaced by the pixels of the EM-CCD. To characterize the confocal volume in this case, we first performed FCS measurements with one central spot and compared the results using for excitation, either the direct path, or the SLM path and for fluorescence detection, either a conventional pinhole and avalanche photodiode (APD), or the EM-CCD.

The reflecting surface of a SLM is generally not perfectly flat, so that the wavefront is slightly distorted after being reflected. To estimate the influence of aberrations introduced by the SLM on FCS measurements, we compared the autocorrelation functions obtained with the two different excitation paths. The direct light path simply includes a beam expander, whereas the SLM path consists of a beam expander, a half-wave plate, a polarizer, the SLM and a collimating lens. When the SLM path is used, the SLM is addressed with the adequate phase map to create only one central spot, so that the signal can be detected using a multimode fiber connected to the APD. The optical setup was designed so that the size of the laser beam at the entrance of the objective lens is unchanged whichever path is taken. It slightly underfills the objective lens back aperture, so that the excitation spot retains a Gaussian profile in the focal plane of the microscope. FCS measurements were performed on a 10 nM-aqueous solution of SRG. The autocorrelation functions (data not shown) exhibit an increase of the average number of molecules N and the residence time (or diffusion time) $\tau_D$ when the laser beam follows the SLM path, which clearly indicate a larger effective volume due to SLM excitation. This tendency was expected since the SLM causes distortions in the wavefront entering the objective lens, so that the focal spot is enlarged. Quantitative estimation of the variation in effective volume can be obtained from the ratio $N_{SLM}/N_{direct}$, whereas the variation of $\tau_D$ accounts for a change in its lateral dimensions. By comparing the number of molecules, we found on average a 20% increase of the effective volume, when using the SLM path. This enlargement seems to be entirely due to an increase of the lateral width of the focal spot (as found from the corresponding increased diffusion time), while its axial length remains unchanged. Therefore, the SLM only slightly degrades the quality of the beam wavefront and its consequences on FCS measurements are reasonably small. However, it should be possible to obtain a smaller excitation spot by using the SLM itself to correct the aberrations of the optical system and its lack of flatness.

Now we turn to the detection side where the fibered APD is replaced by an EM-CCD camera. The first consequence is a loss in time resolution. In this experiment, a subregion of 5☐5 pixels at the center of the CCD chip was read, resulting in a frame rate of 4347.8 Hz (0.23 ms time resolution). Since the time resolution is limited, a sample that exhibits rather slow fluorescence fluctuations is needed. This is why we used a solution of SRG in a 95:5 glycerol-water mixture. Due to the high viscosity of glycerol, molecular motion is drastically slowed down. The diffusion times are typically in the millisecond range (as expected), which is compatible with the frame rate of the EM-CCD camera. However, practically, it is difficult to precisely predict the expected diffusion time of the mixture, because it dramatically depends upon the exact glycerol:water ratio, in addition to its refraction index. Nevertheless, this solution shows the advantage of being very stable in time, compared to solutions of larger objects (micro-beads, for example) which are subject to aggregation problems. Consequently it is well suited to make comparisons between different optical and acquisition systems.

On Figure 3, the autocorrelation function measured with the APD shows two different contributions. The short time contribution is dominated by photophysical effects. The longer time behavior is due to the diffusion of the molecules across the confocal volume. Data fit is performed using the following mathematical model:

$$G(\tau) = 1 + \frac{1}{N} \frac{1 + B\exp(-\tau/\tau_{st})}{\left(1 + \frac{\tau}{\tau_D}\right)\sqrt{1 + \frac{\tau}{S^2 \tau_D}}} \qquad (6)$$

where $N$ is the average molecule number, $B$ is the amplitude of the short-time contribution, $\tau_{st}$ is the effective short-time decay constant, $S$ is the structure parameter ($S=z_0/r_0$ where $z_0$ and $r_0$ are the axial and lateral $1/e^2$ radii of the effective volume) and $\tau_D$ is the diffusion time.

The comparisons between the APD and EM-CCD acquisitions have been done by adjusting the laser power so that the count rate (measured with the APD) is kept constant (this is controlled by removing the mirror sending the fluorescence signal to the EM-CCD, as can be seen in Figure 1). Except for the limited time resolution which makes it impossible to see the faster dynamics using the EM-CCD, both detectors give very similar results when number of molecules and diffusion times are considered. When the direct excitation path is used, one can see that both the number of molecules and the diffusion time are slightly reduced when using the EM-CCD. This might be due to a small difference in effective focal volume between the two detection paths. Although the theoretical magnification of the optical setup should lead to similar detection volumes (the 24 µm square CCD pixel should be equivalent to a 96 µm-square aperture, to be compared to the 100µm-diameter fiber core), the magnification is very sensitive to the exact positions of the lenses and may slightly differ from its theoretical value.

When the SLM excitation and the EM-CCD detection are combined, the autocorrelation function shown on Figure 3 is also close to that of the conventional FCS setup. The number of molecules is slightly higher, since, as mentioned previously, wavefront distortion caused by the SLM leads to an increase of the focal volume. The diffusion time does not seem to increase in proportion. However, as the viscosity of glycerol varies rapidly with the temperature which may have changed slightly when we switched from direct to SLM excitation path, comparison of diffusion times may not be reliable.

In conclusion, this characterization step validates our approach that uses SLM and EM-CCD for parallel confocal excitation and detection. For one central spot in solution, there is no significant deterioration of FCS results: the confocal volume and signal-to-noise ratio are similar to that of conventional FCS measurements.

Further characterization involved measurements with an off-axis spot as a function of the distance to the center. Due to the axial symmetry of the set-up we performed such measurements only in the horizontal direction and we consider that the results obtained would be the same for the other directions. Different phase maps were calculated to create single excitation spots at respectively 0, 1, 2, 4, 6, 8, 10, 12 and 14 µm from the objective axis. The EM-CCD camera was used in the ROI mode and a subregion of 3 lines at the center of the camera was read, so that the Frame Rate was 4545.5 Hz. Lastly, we used the reference solution of SRG at a concentration of about 10 nM in a 95:5 glycerol-water mixture, and we performed for each spot position 25 acquisitions of about 10 s each (45000 frames).

For each spot position, the 25 individual autocorrelation functions were averaged and the corresponding SEM (standard error of the mean) was also computed. The average autocorrelation curve (weighted by the SEM) was fitted by a 3D standard diffusion model, providing values of the number of molecules, $N$, and of the diffusion time, $\tau_D$. For the central spot, *i.e.* at the position 0 µm, we measured an average number of molecules of $N_c = 4.2 \pm 0.2$ and a diffusion time of $\tau_{DC} = 3855 \pm 250$ µs that implies intrinsic variations of about 5 %. If we consider the results for all positions of the excitation spot, the average and standard deviation of the number of molecules and diffusion times are $N = 4.1 \pm 0.3$ and $\tau_D = 3800 \pm 280$ µs respectively. Therefore, the variations observed when changing the spot position are not significantly larger than the uncertainty of the central spot measurements, which means that the confocal volume is almost not deteriorated when the spot is moved off-axis. Moreover, part of the volume variations would be due to misalignments between the excitation spot and the camera pixel which acts as pinhole for the detection. In conclusion, the size of the observation volume stays fairly constant as it is moved across the field of view.

Finally we created several spots to carry out parallel FCS measurements (*i.e.* mFCS).

**4.1.2. Multiple spot measurements**

mFCS measurements were performed on SRG molecules diluted in the 95:5 glycerol-water mixture. By means of the SLM, 7 aligned laser spots were created simultaneously within the solution, as can be seen in Fig.4. The emitted fluorescent light of each volume was focused on one pixel, located in the second lowest row of the EM-CCD camera chip. The distance of the spots from the objective optical axis ranged from 1.2 µm to 17.3 µm, while the distance between neighbouring spots varied from 4.9 µm to 6 µm. The largest difference in the spot intensity was found to be 48%. The reason for this intensity variability can be attributed to either, optical aberrations (SLM and optical path), or to the approximations made for the current phase map calculation (see section 3.2), or to polarisation effects.

In order to accelerate the readout process, the Crop mode was used. In this mode only data coming from the three lowest rows at the bottom of the chip were acquired and thus a frame rate of 9930.5 Hz (*i.e.* a time resolution of $\cong 0.1$ ms) could be achieved.

Two series of measurements, consisting of five acquisitions each, were performed. One acquisition lasted 10 s and consisted of $10^5$ images. The background correction and the calculation of the autocorrelation functions were done as described above for a single spot measurement. For each spot, the 10 autocorrelation functions were averaged and the resulting autocorrelation function was fitted by a 3D normal diffusion model, providing values of the number of molecules, $N$, and of the diffusion time, $\tau_D$ (typical autocorrelation curves and related fits are shown in Fig.4). The obtained values of $N$ and $\tau_D$ in the 7 observation volumes are also shown, together with the background corrected intensity of the corresponding spot. The values of $N$ and $\tau_D$ can be seen to differ from each other. However, the standard error of the diffusion time, $\tau_D$, is the same, whether it is evaluated through the series of 10 acquisitions of a given spot, or throughout the seven different spots (data not shown). Conversely, the number of molecules, $N$, seems to be dependent upon the spot intensity, the highest intensity corresponding to the smallest number of molecules. It must be pointed out that this is not due to real variations of concentration. In fact, the laser power is not uniformly distributed between the different spots, so that the fluorescence background (from neighbouring spots and from zero order diffraction by the SLM) influences mostly the weaker spots, by adding uncorrelated photons that deplete the amplitude of the autocorrelation functions and thus increase the apparent number of molecules. While this effect remains to be further investigated, FCS measurements with multiple spots in solution provide sufficiently reliable results to start applying this technique to living cells.

## 4.2. Measurements in living cells
### 4.2.1. Assessing G actin diffusion

More than 200 measurements have been performed to assess the actin diffusion within about 30 fibroblast cells. In order to achieve the maximum time resolution (*i.e.* $\cong 100$ µs), we chose to align the spots, in the same manner as for multiple spot measurements in solution (see section 4.1.2). The focus was adjusted slightly above the basal membrane of the cells so that the spots were all located within the cytoplasm, avoiding measurement inside the nucleus: a few examples of autocorrelation functions, obtained for a given cell, are shown in Figure 5. Despite our search for a dependency of the autocorrelation functions *versus* the location of the spots close or far from actin fibers, we could not establish such clear correlation. Consequently we only deal with a global analysis. At a first glance, one could think that the FCS study of actin within cells could be described by some simple diffusion-interaction laws, analogously to what has already been done with FRAP or photoactivation experiments (31). If this was true, a model such as the one of Michelman-Ribeiro *et al.* (36) would provide independent estimations of the $k_{on}$, $k_{off}$ attachment and detachment rates with some immobile structure, together with the diffusion constant of the free particles. Unfortunately this is not possible for several reasons. The dynamics of actin is governed by interactions with a complex network of proteins among which, not only are monomeric (or globular) actin, G-actin and filamentous actin, but also numerous regulators of actin assembly. The actin filament turnover characteristic time (several minutes) is far beyond the time range of FCS. The detachment rates at the pointed and barbed filament ends could, in principle, be assessed by FCS, since one expects some of the corresponding times to be as small as 0.2 s (30). However, practically, global variations of the fluorescence signal on similar time scales (due to cell motility, deformation, *etc.*) would also affect the signal and, consequently, preclude any reliable estimation of $k_{off}$. Consequently, we have used an effective two component model (see Eq. (5) in 3.3.2) to fit the data and found a value of $977 \pm 414$ µs for the short diffusion time, which is believed to be much more reliable than the long time. One sees in Figure 6a the corresponding histogram obtained from an ensemble of 200 fits. To estimate the diffusion constant from this value of diffusion time, we assume the diffusion constant of SRG in pure water to be similar to that of rhodamine 6G, *i.e.* around 400 $\mu m^2/s$ (37). This corresponds to a diffusion time of $\cong 30$ µs, when measured by using the fast APD detection. Since the present mFCS system gives diffusion times close to those of the APD detection (see Figure 3), the mean value of 977 µs corresponds to a diffusion constant of about 12 $\mu m^2/s$, which is in good agreement with the 15 $\mu m^2/s$ value measured by FRAP (37). The longer effective diffusion time, see Figure 6b, probably reflects the complex network of interacting molecules, but might also be biased by global variations of the fluorescence intensity.

### 4.2.2. Response to heat shock

In order to study the response to heat shock by using mFCS, we have located 6 spots throughout the whole nucleus, using the multi-track mode, with the consequence that the acquisition rate is lower by a factor 2 compared to the Crop mode. Thus, the time resolution of the autocorrelation curves is now limited to about 200 µs. Nevertheless, it clearly permits to see differences induced by the heat shock: as can be seen in Figure 7a, the number of HSF1 molecules decreases, on average, when heat shocking, while the dynamics is modified (Figure 7b). The large error bars associated with the amplitude of the autocorrelation curve (Figure 7a) put into evidence the variability of the molecule concentration that contrasts with the much more reproducible shape of the autocorrelation curves, as seen in Figure 7b.

The decrease of the HSF1 concentration in the nucleoplasm outside the nuclear stress bodies corresponds in the fast relocalization of HSF1 within the NSBs (33). The slower dynamics reflects the HSF1 trimerization, DNA binding to the promoter sites (HSE) of the heat shock proteins (HSP) and protein interactions that occurs upon exposure to stress conditions (33, 39). This preliminary study shows that our system allows us to induce heat shock and to perform dynamics measurements in various locations and times.

## 5. DISCUSSION AND CONCLUSION

The results presented in this paper are to be considered as a proof of principle of the mFCS technique. We have shown that the spatial resolution is very close to that of a standard FCS set-up (*i.e.* with a usual pinhole and APD detection). More precisely, the number of molecules and diffusion time obtained by combining the SLM excitation with the EM-CCD detection are close to those of the direct excitation and APD detection. This also demonstrates that the optical aberrations introduced by the SLM are limited. We have developed an algorithm that computes the phase map addressed to the SLM, in order to generate a series of laser spots at different places of the field of view. However, the homogeneity of the spot intensities has to be improved. There are several possible reasons (under studies) for this inhomogeneity, such as the approximations used in the computation of the phase map addressed to the SLM, the polarization of the outgoing beam, or the residual optical aberrations introduced by the whole optical bench. Additionally, an iterative algorithm similar to the Gerchberg-Saxton algorithm (35) could help to improve the homogeneity of intensities.

The multiplexed FCS acquisition can be performed using an EM-CCD camera, the price to pay being a reduced time resolution (≅ 100 µs), compared to usual FCS devices using APD. We are currently investigating ways to improve the time resolution (down to a few 10 µs), using well suited readout modes of the EM-CCD camera. Fast cellular processes, like signal transduction, would benefit from such improvements. Nevertheless the current time resolution is often sufficient for living cell measurements. As a matter of fact we have been able to measure the G-actin diffusion constant (which, to our knowledge, had never been measured by FCS), thus confirming values previously measured by FRAP (31, 38). Another encouraging application of the mFCS technique is the study of the response to heat shock, where dramatic spatial and temporal variations of HSF1 concentration must occur. We have observed the modification of the shape of the autocorrelation curves during heat shock, together with the number of molecules, which reflects both relocalization of HSF1 and trimerization. In addition, EM-CCD rather than APD detector makes it possible to measure highly fluorescent structures, such as nuclear stress bodies thereby giving access to DNA binding dynamics.

A challenging application of mFCS would be to measure active transport, since, by performing cross-correlation between the fluorescence signals of different spots, one can access to the molecular flow (25, 27). Another interesting extension of the present mFCS set-up would be a multi FRAP device (40), especially well adapted for the study of slow processes, such as those occurring in the dynamics of the actin cytoskeleton. Since the relevant time scales range from that of FCS (G-actin diffusion) to that of FRAP (actin detachment rates at filament ends and turnover time), an integrated experimental device that would permit to perform both kinds of measurements would be of great interest. Compared to the Raster Image Correlation Spectroscopy (6) or scanning FCS approaches (7), mFCS has the advantage of disentangling the spatial and temporal measurements. Another alternative to mFCS is pinhole array correlation imaging (22). In this case, hundreds of laser spots are systematically generated, covering the whole sample, thus inducing possible drawbacks such as photo-damage and photobleaching. This is especially unfortunate when the fast kinetic readout mode of the camera is to be used, because it needs that a part of the camera is masked (26), as in the CROP mode, while the whole sample is illuminated.

In conclusion, we believe that the present mFCS technique has unique capabilities and, although improvements have to be brought regarding the homogeneity of intensities and time resolution, it is an important step towards spatially resolved multiplexed FCS.

## 6. AKNOWLEDGEMENTS

This project was funded by the French Agence Nationale de la Recherche under contract ANR-08-PCVI-0004-01 and by the Region Rhone-Alpes (CIBLE 2009).

**Abbreviations:** FCS (Fluorescence Correlation Spectroscopy), SLM (Spatial Light Modulator), EM-CCD (Electron Multiplied Charge Coupled Device), ROI (Region Of Interest), APD (Avalanche PhotoDiode), FRAP (Fluorescence Recovery After Photobleaching), PAF (PhotoActivation of Fluorescence)

**Key Words:** Fluorescence Correlation Spectroscopy, microscopy, spatial light modulator, diffusion, single molecules, multi-confocal, actin, cytoskeleton, heat shock

**Send correspondence to:** Antoine Delon, Universite de Grenoble 1, CNRS, Laboratoire de Spectrometrie Physique UMR 5588, BP 87, 38402 Saint Martin d'Heres, France, Tel: 33-0476635801, Fax, 33-0-476635495, E-mail: adelon@ujf-grenoble.fr

**Running title:** Multi-confocal fluorescence correlation spectroscopy

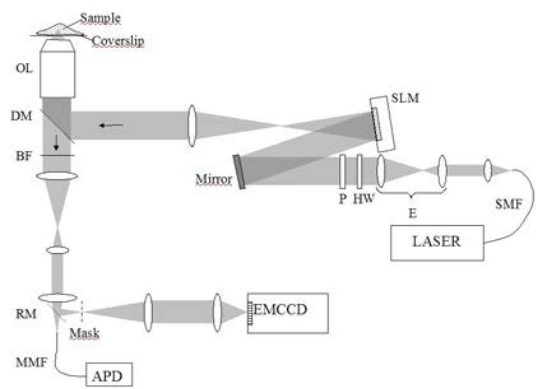

**Figure 1.** Optical setup to perform multiconfocal FCS. Abbreviations: OL, objective lens; DM, dichroic mirror; BF, bandpass filter; RM, removable mirror; P, polarizer; HW, half-wave plate; E, ×3-beam expander; SMF, single-mode fiber; MMF, multi-mode fiber; APD, avalanche photodiode.

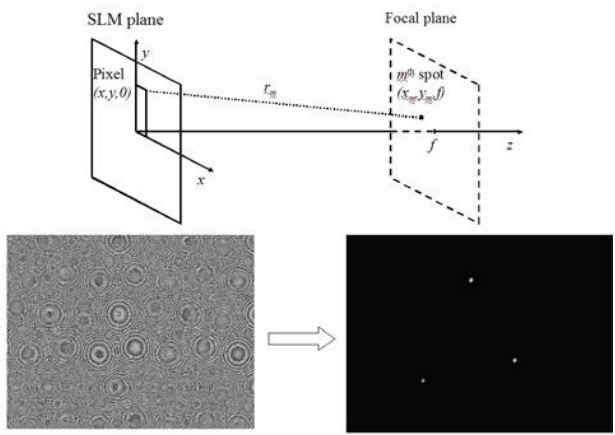

**Figure 2.** Computation of the phase map displayed on the SLM to create multiple spots: upper half: geometry of the pixel and spot positions; lower half: phase function addressed to the SLM to create three spots and the corresponding intensity distribution.

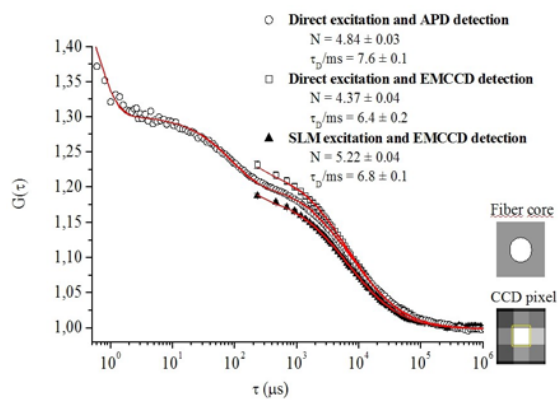

**Figure 3.** Comparison of the autocorrelation functions obtained with: direct excitation path and APD (standard configuration), empty circles; direct excitation path and EM-CCD detection, empty squares; SLM excitation path and EM-CCD detection, solid triangles. Error bars are shown for EM-CCD measurements calculated over 20 acquisitions of 10 seconds each (with a photon count rate of 20 kHz). Data are fitted with Eq. 6 (lines). The average number of molecules $N$ and diffusion times $\tau_D$ obtained by the fit are indicated for each curve. Parameters related to short-time processes which cannot be estimated from the EM-CCD curves were set to the values obtained from fitting the APD curve.

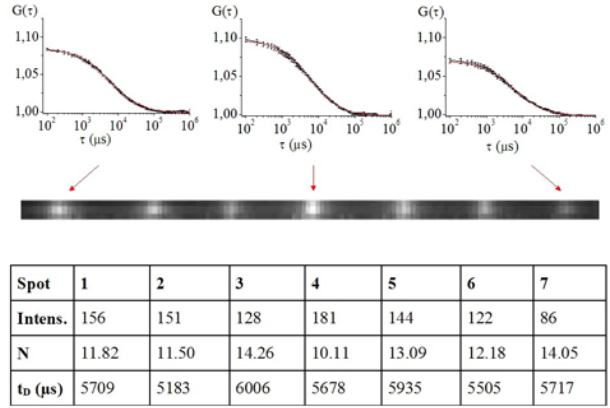

**Figure 4.** Averaged image (over $10^5$ frames) of a series of 7 laser spots aligned on the second lowest row of the CCD chip. Three examples of autocorrelation functions are displayed (corresponding to a middle, low and high intensity spot). The table displays the intensities (background corrected), number of molecules ($N$) and diffusion time ($\tau_D$) of the corresponding spots.

| Spot | 1 | 2 | 3 | 4 | 5 | 6 | 7 |
|---|---|---|---|---|---|---|---|
| Intens. | 156 | 151 | 128 | 181 | 144 | 122 | 86 |
| N | 11.82 | 11.50 | 14.26 | 10.11 | 13.09 | 12.18 | 14.05 |
| $t_D$ (µs) | 5709 | 5183 | 6006 | 5678 | 5935 | 5505 | 5717 |

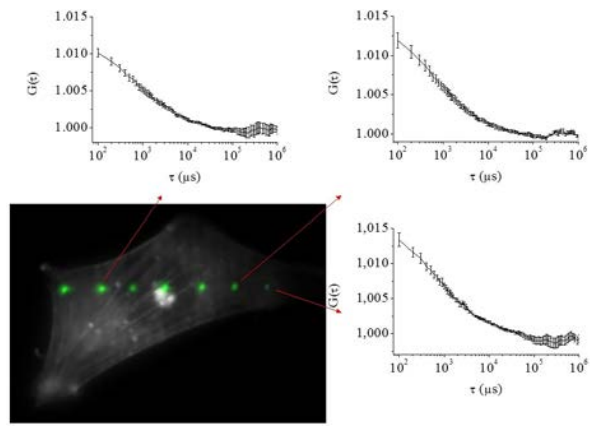

**Figure 5.** An example of an actin-eGFP cell (wide field fluorescence image obtained by flattening the SLM), where the series of 7 laser spots have been superimposed (using the ImageJ software). Among these spots, three autocorrelation functions are displayed.

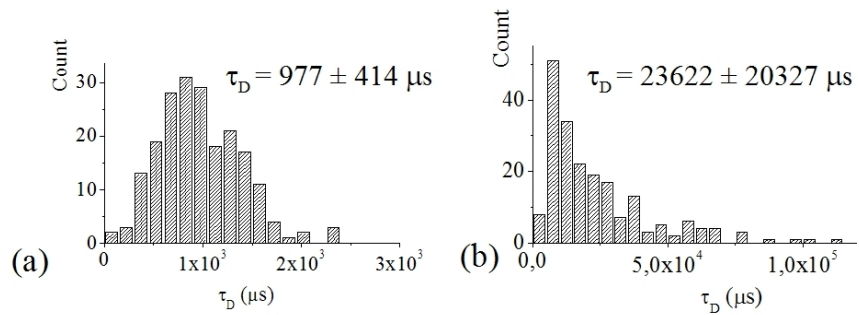

**Figure 6.** Histograms of the two diffusion times obtained by fitting about 200 autocorrelation curves from actin-eGFP cells, with an effective two component model; (a) short diffusion time; (b) long diffusion time.

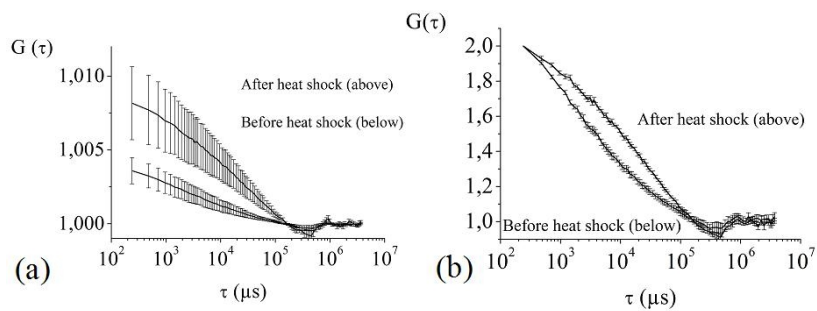

**Figure 7.** Autocorrelation curves obtained before and after heat shocking, by averaging over 26 measurements; (a) without any normalization of the amplitude, thus exemplifying the mean value and dispersion of the number of molecules; (b), with normalization of the amplitudes at $\tau = 240$ µs.